\documentclass[preprint,12pt, authoryear]{elsarticle}
\usepackage{float}
	\usepackage{amsmath}
\usepackage{amssymb}
\usepackage{graphics}
	\graphicspath{{figures/}}
		\usepackage{numprint}
\usepackage{multirow}
\usepackage{arydshln}
\usepackage{placeins}
\usepackage{url}
	\usepackage{lscape}
	\usepackage{pgfgantt}
	\usepackage{subcaption}
	\usepackage{tikz}
\usetikzlibrary{shapes,arrows,decorations}

\usepackage[font=tiny,labelfont=bf]{caption}
\usepackage{chngcntr}

\newcommand{\red}[1]{{\color{black}#1}}
\newcommand{\blue}[1]{{\color{black}#1}}

\usepackage{amssymb}

\usepackage{atbegshi,picture}
\usepackage{lipsum}

\journal{ArXiv}

\begin{document}

\begin{frontmatter}



\title{Do diverse and inclusive workplaces benefit investors? \\ An Empirical Analysis on Europe and the United States}

\author[1,2]{Karoline Bax}

\address[1]{Department of Economics and Management, University of Trento, Via Inama 5, 38122 Trento, Italy}

 \address[2]{Center for Digital Transformation, Technical University of Munich, Bildungscampus 9, 74076 Heilbronn, Germany}


\begin{abstract}
As the COVID-19 pandemic restrictions slow down,  employees start to return to their offices. Hence, the discussions on optimal workplaces and issues of diversity and inclusion have peaked. Previous research has shown that employees and companies benefit from positive workplace changes. This research questions whether allowing for diversity and inclusion criteria in portfolio construction is beneficial to investors. By considering the new Diversity \& Inclusion (D\&I) score by Refinitiv, \blue{I find evidence that investors might suffer lower returns and pay for investing in \textit{responsible} (i.e., more diverse and inclusive) employers in both the US and European market.}
\end{abstract}


\end{frontmatter}



\section{Introduction}	

\red{As infection rates are mostly static now,  and the COVID-19 pandemic reaches an endemic stage\footnote{\blue{see e.g., https://www.mckinsey.com/industries/healthcare-systems-and-services/our-insights/when-will-the-covid-19-pandemic-end}},  employees are slowly returning to their offices.  Thus,  the debate around workplace design and contentment has received increased attention.

}

While companies realized that diverse and inclusive environments create financial benefits,  increase employee satisfaction, and build a productive working environment already prior to the COVID-19 pandemic\footnote{These beneficial effects can take place either directly  \citep{krapivin2018google, Refinitiv2022covid},  indirectly through customer satisfaction  \citep{chi2009employee}, or in an investment setting \citep{mishra2018happy}, and also help the organization to grow  \citep{spreitzer2012creating}.},   the discussion has now gained further traction. 

More specifically, the discussion is subject on the newly introduced Diversity \& Inclusion (D\&I)  score by Refinitiv (formerly ASSET4) \citep{Refinitiv2022a}.  
The score is built using 24 metrics defined on four main pillars linked to workplace well-being \citep{Refinitiv2022a}. These are: diversity (e.g.,  gender diversity and diversity process), inclusion (e.g., flexible working hours and daycare services), people development (e.g., internal promotion, employee satisfaction, and career development process), and news and controversies (e.g., diversity and opportunity and wages and working conditions) \citep{Refinitiv2022a}. The input data is gathered from publicly available sources, and an overall D\&I score is computed. The D\&I  score ranges from 0 to 100, 0 indicating poor D\&I performance.  According to \cite{Refinitiv2022a}, this score shall provide insights into two main issues employers face: cultural diversity and workplace transition. 

Different scholars have already shown that diversity can positively affect economic performance and firm value.\footnote{See for diversity in general in the US \citep{erhardt2003board},  diversity on boards in Spain \citep{ campbell2008gender,reguera2017does} and in Malaysia \citep{lee2017gender}),  diversity on top management teams in Denmark \citep{opstrup2015right}). } Others find mixed effects using a sample of Bangladesh companies  \citep{dutta2006gender} and Italian companies \citep{gordini2017gender}.  Additionally, diversity and inclusion have been linked to workplace happiness,  productivity, and better performance in challenging situations \citep{vasa2018green, arenas2017shaping,frey2010happiness,warr2011work}.  

While this topic has been explored from a company management and human resource perspective,  happiness and satisfaction can technically translate into higher financial performance measures through different channels  \citep{kessler2020job,Refinitiv2022covid},  including increased productivity \citep{utami2018relationship} or employee well-being  \citep{krekel2019employee}. However, measuring and predicting workplace happiness can be complex and different measures and dimensions have been proposed.\footnote{See \cite{fisher2010happiness, sirota2013enthusiastic, wesarat2014conceptual,mousa2021does,arslan2021diversity,barak2002outside}.} For investors to whom the company might be a black box, gathering this kind of information is even more challenging. This is where the new D\&I score could aid as a proxy for workplace happiness and diversity and become a first step in solving the measuring issue for investors \citep{Refinitiv2022a}. 

While it is clear that a diverse and inclusive workplace setting benefits the employees and helps the company to grow, in this work I question whether this setting can also directly benefit investors. More specifically,  as the differences in the welfare social systems in Europe (EU) and in American (US) are so prominent (see e.g., \cite{alber2010european,vaughan2003eu,wickham2002end}), I am focusing on comparing these two markets.  \red{Using a market capitalization weighted portfolio, I compare the out-of-sample performance of portfolios with different mean D\&I scores and identify little to no difference between markets but consistent difference between companies with high versus low D\&I scores for both markets. More specifically,  using a Long-Short portfolio (taking a long position on a portfolio with high D\&I scores and a short position on a portfolio with low D\&I scores), I find evidence that investors do not benefit from additional gains and might have to suffer from lower returns when investing into companies with \textit{responsible}  employers (i.e., companies with high D\&I scores). }This is evident in the US and in the European market. 

The study proceeds as follows: Section 2 describes the methodology and the data. Section 3 elaborates results, and Section 4 concludes the study.

\section{Methodology and data}
I consider the most recent D\&I scores, return data, and market capitalization weight data of 363 EU companies which are constituents of the EURO STOXX 600 (STOXX) index, and 365 US companies which are  constituents of the Standard \& Poor’ (S\&P) 500 index (SPX) and whose observations are continuously available from January 3, \blue{2020 to August, 31 2022}. \blue{The return data, market capitalization,  and the D\&I scores were retrieved on September 16, 2022 from Refinitiv (formerly ASSET4).} \blue{The time period was chosen so that it includes the COVID-19 outbreak, the pandemic, and now the endemic stage.} The reason for not including all constituents of these indices is that the D\&I score is not available for every single \blue{asset.} While I include assets that have a zero score,  I exclude assets that have no associated D\&I score.\footnote{Indicated by Refinitiv as \textit{NA}.}
 \blue{ The datasets are similar in their descriptive statistics; nevertheless, the D\&I scores of the constituents of the SPX show larger kurtosis and skewness as reported in the Tables \ref{des} and \ref{desDI}.}

\begin{table}[H]
\centering
\tiny
\npdecimalsign{.}
\nprounddigits{2}
\begin{tabular}{|c|c|c|n{2}{5}|n{2}{5}|n{2}{5}|n{2}{5}|c|c|c|}
  \hline
 \textsc{Dataset} &  \textsc{\hspace{0.2cm}$T$}  &  \textsc{$N$}  & \textsc{\hspace{0.5cm}$\hat{\mu}$}&  \textsc{\hspace{0.3cm}$\hat{\sigma}$}  &\textsc{$\widehat{skew}$}& \textsc{$\widehat{kurt}$}& \textsc{$period$} &  \textsc{$freq.$}  \\ 
  \hline
 \textsc{SPX} & 672  & 363 &0.13447 & 0.17078 & -0.071548 & 13.8101 &  \textsc{03/01/2020 - 31/08/2022} &  \textsc{daily}  \\
  \textsc{STOXX} &   686 & 365 &0.038358 & 0.14232 & -0.25498 & 15.584 & \textsc{03/01/2020 - 31/08/2022 } &  \textsc{daily} \\ 
   \hline
\end{tabular}
\caption{The table reports the descriptive statistics for the asset returns of the constituents of the Eurostoxx 600 and the S\&P 100 (SP100), respectively.  Columns 1-9 report the number of observations ($T$),  the number of constituents ($k$), the average annualized mean ($\hat{\mu}$), the average annualized standard deviation ($\hat{\sigma}$), the average skewness ($\widehat{skew}$), the average kurtosis ($\widehat{kurt}$) of the asset returns,  the time period ($period$) and the sampling  frequency ($freq.$).} 
\label{des}
\end{table}

\begin{table}[H] 
\centering
\small
\begin{tabular}{|l|llll|}
  \hline
 \textsc{Dataset} & $\hat{\mu}$& $\hat{\sigma}$& $\widehat{skew}$ & $\widehat{kurt}$\\ 
\hline
SPX&50.24 & 450.13& -1.56 & 4.36 \\ 
STOXX& 48.46& 652.03 & -1.19 & 2.84 \\ 
\hline 
\end{tabular}
\caption{The table reports the descriptive statistics for the D\&I score of the constituents of the Eurostoxx 600 and the S\&P 100 (SPX), respectively.  Columns 1-4 report the the average mean ($\hat{\mu}$), the average standard deviation ($\hat{\sigma}$), the average skewness ($\widehat{skew}$), the average kurtosis ($\widehat{kurt}$) of the D\&I scores.} 
\label{desDI}
\end{table}

 In both datasets, there are some assets (\blue{48 assets for SPX and 74 for STOXX}) that have a score equal to zero; the remaining assets have a score in higher quartiles. According to \cite{Refinitiv2022b}, companies are only given a non-zero D\&I score if they have non-zero scores on all four pillars. This means that companies with a zero score might, on some level, contribute to diversity and inclusion; however, not on all dimensions leading to a zero score. Furthermore,  D\&I scores are benchmarked against the country.\footnote{Examples here are the sub metrics of the diversity pillar: \textit{Percentage of females on the board} and the inclusion pillar: \textit{Percentage of employees with disabilities or special needs} \cite{Refinitiv2022b}.} Therefore, employee benefits and diversity and inclusion efforts are not directly comparable across countries. However, the scores allow for an in-country comparison. More specifically,
a high D\&I score of a US company indicates that this company is performing better than most others in the USA in terms of diversity and inclusion. The same line of argumentation can be made for a company in Europe. However, if one directly compares both scores of US and EU companies, it is problematic as the score indicates the relative performance within the respective origin. 
Due to the limitation of the scores as well as the differences in the welfare social systems in the EU versus the US, individual portfolios for the EU and US are created and their performance is compared.

I use the most recent D\&I score to categorize the assets into different portfolios for each dataset, as also suggested recently for ESG scores by \cite{cerqueti2021esg}.\footnote{\blue{One reviewer pointed out that it would be more accurate to re-balance the portfolio periodically according to the corresponding D\&I score.  This is true and can be applied in future research, however, the data is right now not available from Refinitiv. }} More specifically, to compare the effect of D\&I scores on investors, \blue{four portfolios (from now on referred to as PT) are formed using evenly spaced cumulative probabilities to divide the assets into quartiles using the D\&I score. This approach is similar to ESG portfolios, as assets can be attributed to a rating class (i.e., $A,B,C$, or $D$) based on its ESG score value using thresholds or quartiles.  Using quartiles instead of thresholds allows to create similarly sized portfolios, making performance comparison more accurate.  PT 1 contains the assets with the lowest  D\&I score while PT 4 contains the assets with the highest D\&I score. \blue{The number of assets within the portfolios is equal to PT 1-4 = 91,	92,	93,	89 assets for the US portfolios and PT 1-4 = 91,	96, 86, 90 assets for the EU portfolios. }
 }

\blue{The process of dividing the assets into different portfolios based on their D\&I score is similar to the most prominent approach in ESG investing.  According to \cite{amel2018and}, most investors use the ESG score to follow a  negative screening approach.  More specifically, assets with a low ESG score are excluded from the investment portfolio.   The rational is that investors tend to believe that additional non-financial information is an indicator for different company risks, including reputational, legal,  and regulatory risk \citep{amel2018and}. Also, \cite{verheyden2016esg} showed that there is a positive effect of ESG screening resulting in improved risk-adjusted returns. As D\&I scores also provide additional non-financial information to the investor, a similar approach is used here.

 }

\subsection{Empirical Set-up and Performance Measures}

To examine the effect of different D\&I scores for investors,  I use the previously created portfolios defined by grouping the assets into quartiles according to their D\&I score (PT 1-PT 4 US and PT 1-PT 4  EU) \red{and analyse their out-of-sample returns with a  factor model.  }

\red{The weights of the assets in the portfolios are determined by the market capitalization of the assets.}\footnote{\red{Other portfolio construction strategies (e.g.,  Equally-Weighted or Minimum-Variance) have resulted in a similar outcome.}} The resulting portfolio is assumed to be held for one day. 
Then, I roll the look-back window forward by one day. By doing so, I discard the oldest observation and include one new observation. This process is repeated until the end of the time series is reached. Using $M$ out-of-sample returns, where  ${r}_{t+1}$ is defined as the vector of returns at time $t+1$, I evaluate the resulting portfolios in terms of risk and return profile and portfolio composition.
 \blue{Following this process, I generate for window size $ws=50$, $M=635$ out-of-sample returns for the EU portfolios and $M=621$ out-of-sample returns for the US portfolios.}\footnote{\blue{The analysis has been run additionally on $ws=25,84,170$ and results are similar.}}

\red{
To test the performance of the four portfolios, I run a factor model on the risk-adjusted out-of-sample portfolio returns to calculate the risk-adjusted abnormal performance of the four market weighted portfolios.\footnote{\red{The analysis has been repeated using the CAPM, Fama-French 3 - Factor model, and Cahart model and found qualitatively similar results.}} This set up is inspired by  \cite{pavlova2022esg}.  More specifically, I am using the  Fama-French 5-factor model plus the momentum factor.\footnote{\blue{Data used to estimate these models was retrieved on 21. September 2022  from https://mba.tuck.dartmouth.edu/.}} \blue{Thus, I get}

\begin{equation}
\begin{aligned}
R_t-R_{f t}=&\alpha+\beta_1\left(R_{m t}-R_{f t}\right)+\beta_2\left(S M B_t\right)+\beta_3\left(H M L_t\right)\\ &+\beta_4\left(R M W_t\right)+\beta_5\left(C M A_t\right)+\beta_6\left(MOM_t\right)+\varepsilon_t
\end{aligned}
\end{equation}
}

\blue{

where $R_t$ is defined as the portfolio return at time $t$ for the four different portfolios of the market weighted portfolio construction.

Then, $R_{m t}-R_{f t}$ is  defined as the the excess return on the market, $R_{f t}$ is the risk-free rate, $S M B_t$ and $H M L_t$ are the size and value factors, respectively of  3-factor model by \cite{fama1993common},  while $R M W_t$ and $C M A_t$ are the profitability and investment factors  of the \cite{famafrenchfive} 5-factor model.  \red{While the first factors capture the small minus big and high to low book-to-market ratio, the last two factors measure the return spread of the most profitable companies minus the least profitable companies as well as  conservatively minus aggressively investing.  Then, the $MOM_t$ denotes the momentum factor and aims to capture the tendency of a company to do well \citep{jegadeesh1993returns}. Lastly,  $\varepsilon_t$ is the error term of the model.}}
\red{By estimating this factor model,  a closer look at the alpha can be taken.  This allows me to get a preliminary understanding of the performance of the portfolios with different D\&I scores.  However, this approach lacks comparability across portfolios.

Therefore, an additional approach allowing to compare the top D\&I performing companies versus the bottom D\&I performing companies has been implemented and strengthens the analysis.  In order to be able to directly compare these portfolios a Long-Short strategy has been executed.  More specifically,  a long position is taken on the higher D\&I score companies (PT 4) and a short position on the companies with the lowest D\&I score (PT 1).   Then, by running the same factor model as described in Equation 1,  the alpha of the Long-Short portfolio (PT 4 - PT 1) is analyzed. The reason for setting up this portfolio strategy is to analyze whether the top D\&I score portfolio can provide an additional gain after considering the most common risk factors.  Thus, I anticipate the PT 4 would increase in value while PT 1 would decrease. Then, if the alpha of the Long-Short portfolio (LS PT) is positive and statistically significant,  investing into D\&I scores can provide an additional gain for investors.
}

\FloatBarrier

\section{Empirical Results}
\blue{ The out-of-sample return performance using the factor models for the five (PT 1-PT 4 and LS PT) market weighted portfolios is reported in Table \ref{MCSPXSTOXX}.
\red{
Starting with the PT 1 which includes the assets with the lowest D\&I scores,  the alphas are significant and  positive for both markets. This is quite different to the portfolios with higher D\&I scores,  namely PT 2 to PT 4. The significant alphas for these groups are negative. Also comparing US versus EU, I find little difference in the magnitude of  the alphas for PT 3 and PT 4.  

For both,  the US and EU portfolios,  these results suggest that investor who take higher D\&I risk (invest into companies with lower D\&I scores) may be rewarded with higher returns.  Portfolios with higher D\&I scores (PT 3 and PT 4) seem to underperform,  i.e.,  alphas are negative and mostly significant.

After this preliminary evidence,  a closer look at the Long-Short portfolio is taken.  Both alphas are negative, while the result for the US market (SPX) is statistically significant.  This indicates that investors cannot benefit from additional gains after controlling for the additional risk factors when investing into companies with high D\&I scores.  More specifically,  the portfolio is under performing and no excess return is generated indicating the higher D\&I scores of companies are not a beneficial investment criteria. }

\begin{table}[hbt!]
\centering
\begin{tabular}{lll}
\hline
Portfolios   & FF5  + MOM $\alpha$ & FF5  + MOM $\alpha$ \\
\hline
Market       & SPX                 & STOXX               \\ \hline
Portfolio 1  & 0.0005***           & 0.0005***           \\
Portfolio 2  & -0.0197             & -0.0419             \\
Portfolio 3  & -0.0006             & -0.0005**           \\
Portfolio 4  & -0.0005***          & -0.0005***          \\ 
LS Portfolio &                  -0.003**   &                  -0.0002  \\
\hline
\end{tabular}
\caption{Market Weighted Portfolio using the SPX data (column 2)  and STOXX data (column 3) with $ws=50$. Other $ws$ are similar and available upon request from the author.*,**, *** p-values indicate significance at the 10\%, 5\%, and 1\% level, respectively. }
\label{MCSPXSTOXX}
\end{table}

\FloatBarrier

}

\FloatBarrier

\blue{From the empirical results, one can argue that investors have to pay more in order to invest 
into more diverse and inclusive companies.  This is similar to the idea that investors have to pay to be green and higher
 ESG risks can be rewarded with higher returns  (see i.e.,  \cite{pavlova2022esg,pastor2020mutual,pyles2020examining}), 
this concept could also apply for the D\&I scores.   
\red{One possible reason for these findings is the extensive effort companies have to take in order to fall into the high D\&I score category.  These include for example offering Day Care services and a large number of skill and career development trainings \citep{Refinitiv2022a}.  Additionally,  wages and working conditions are also also critically assessed when the D\&I scores are computed \citep{Refinitiv2022a}. All of these variables mentioned,   result in additional costs for the companies.  Thus,  they possibly negatively impact the overall financial performance of the companies.}
Nevertheless, the request for changes and flexibility have only recently 
gained momentum,  and possibly in the future investors might be able to benefit from investing into companies with \textit{responsible}  
employers.  \red{Technically,  these companies should attract higher quality employees which in the end might be more productive and lead to a more qualified workforce.}

Additionally, the need for periodic D\&I scores is prominent and could also improve the analysis further.
 }

 \FloatBarrier

\section{Conclusion}
To conclude, in this new COVID-influenced world, diverse and inclusive work environments and employee satisfaction have gained importance. While research has already shown large benefits for employees and companies in general when adopting more diverse and inclusive strategies, in this paper, I question whether a diverse and inclusive workplace should be taken into account by investors. Acting on this information became now significantly easier as Refinitiv's new D\&I score allows investors to easily gather information on the workplace culture of many different companies.

By creating portfolios with different mean D\&I scores, \blue{I find evidence that abnormal returns are generally only notable for portfolios of lower-scoring (i.e., low D\&I score) companies, while companies with high D\&I scores tend to show no abnormal returns for investors. \red{This finding is also supported when using a Long-Short portfolio, taking a long position on the portfolio with high D\&I scores and a short position on the portfolio with low D\&I scores.  The negative alphas for both markets indicate that this strategy does not generate  additional gains after controlling for relevant risk factors. 

At this point, I conclude that investors cannot benefit from superior financial performance when investing into companies with \textit{responsible} employers.  }Regulators should take these findings as  incentives to foster change in order to steer investments into  \textit{responsible} employment which then hopefully benefits the investors as well as the employees. }

\blue{These findings are, however, limited by the data availability of the D\&I score. There is a pressing need for periodic D\&I scores which would make the analysis more accurate.  Moreover,  the paper could be updated in the future when more data providers have published D\&I scores. }Furthermore, building a time series for D\&I indicators is high up on the agenda.   

Future research could look into differences within Europe and consider different periods to distinguish the effect in times of crisis. Additionally, it could be interesting to consider the time period prior to, during, and ideally after COVID-19 is no longer a  threat.

\section*{Acknowledgements}
I am grateful to the referee and editor for commenting and
leading to a considerably improved version. 

  \bibliographystyle{elsarticle-harv} 
  \bibliography{References/ref_diversity.bib}

\end{document}